\documentclass[11pt]{article}
\usepackage{charter}
\usepackage{amssymb,amsmath,graphicx,url}
\usepackage{setspace,upgreek,amsmath}
\usepackage{geometry}[margin=0.75in]
\usepackage{lineno}

\usepackage[sort&compress,numbers,comma,square]{natbib}

\title{Why multiple hypothesis test corrections provide poor control of false positives in the real world}
\author{Stanley E. Lazic\textsuperscript{1,*}}
\date{}

\begin{document}
\maketitle

\textsuperscript{1}\emph{Prioris.ai Inc., 459--207 Bank Street, Ottawa,
K2P 2N2, Canada}

\textsuperscript{*}Corresponding author: \texttt{stan.lazic@cantab.net}

\section*{Abstract}
Most scientific disciplines use significance testing to draw conclusions about experimental or observational data. This classical approach provides a theoretical guarantee for controlling the number of false positives across a set of hypothesis tests, making it an appealing framework for scientists seeking to limit the number of false effects or associations that they claim to observe. Unfortunately, this theoretical guarantee applies to few experiments, and the true false positive rate (FPR) is much higher. Scientists have plenty of freedom to choose the error rate to control, the tests to include in the adjustment, and the method of correction, making strong error control difficult to attain. In addition, hypotheses are often tested after finding unexpected relationships or patterns, the data are analysed in several ways, and analyses may be run repeatedly as data accumulate. As a result, adjusted p-values are too small, incorrect conclusions are often reached, and results are harder to reproduce. In the following, I argue why the FPR is rarely controlled meaningfully and why shrinking parameter estimates is preferable to p-value adjustments.

\section*{Introduction}

Scientists are often taught to analyse data using a null hypothesis significance testing (NHST) framework. This involves posing the research question as two mutually exclusive hypotheses. The first is the null hypothesis ($H_0$), which asserts that there is no effect, correlation, or association. The second is the alternative hypothesis ($H_1$), which states there is an effect, correlation, or association, or more simply, that $H_0$ is false. The data are then tested for compatibility with the null hypothesis by reducing them to a single number, called the test statistic, which is then used to calculate a p-value. A small p-value indicates that the observed or a more extreme test statistic is unlikely if the null hypothesis is true (and all assumptions of the test are met, and there are no unmeasured biases). Informally, a small p-value indicates that the observed effect, correlation, or association is larger than expected, if nothing is actually occurring. Based on this argument, the alternative hypothesis must be more probable because there are only two alternatives ($H_0$ and $H_1$), and if one is discredited, the other must be more likely. This implication does not follow directly since a p-value is the probability of data, not a hypothesis. Nevertheless, researchers often interpret a p-value as the probability that a hypothesis is correct. NHST's practical and philosophical limitations are not considered here, as they have been covered elsewhere. \cite{Cohen1994,Goodman2008,Stang2010}.

A false positive, also known as a Type I error, occurs when no effect is present ($H_0$ is true), but a researcher incorrectly concludes that an effect exists. The false positive rate is the number of false positives in a collection of tests. When using NHST, a predetermined threshold is set to control the false positive error rate (denoted $\alpha$ and often set to 0.05), and if the p-value is less than $\alpha$, $H_0$ is rejected. When $\alpha = 0.05$, there is a 5\% chance that an incorrect conclusion will be reached, by definition, assuming $H_0$ is true. If many such tests are conducted, there is a greater than 5\% chance that at \emph{least one} test in the collection will be a false positive. In response to this increased error rate, post-hoc tests or multiple comparison procedures (MCP) have been developed, which continues to this day. I argue that the MCP framework does not control the false positive rate (FPR) in many real studies, and therefore contributes to the poor reproducibility of research.

Researchers have proposed several reasons for irreproducibility, including publication bias, selective reporting, inappropriate experimental designs and analyses, and perverse incentives \cite{Landis2012,Ioannidis2014a,Begley2015,Lazic2016}. A more basic explanation has been overlooked: the true FPR is far from that assumed in many experiments, even when analyses are conducted appropriately. I describe below why the FPR is rarely controlled in any meaningful way in many experiments. Some reasons are well known and have good solutions, but others are unrecognised. The NHST approach also assumes that there is a single unique $\alpha$ to control, but I argue that this is a myth. This paper concludes with suggestions for improving inferences when testing multiple hypotheses.

\section*{Reasons for poor control of false positives}

Many experiments will suffer from at least one of the nine reasons given below for why the FPR may differ from the nominal 0.05 level. These items describe current practices in biology, which vary from field to field, and do not necessarily reflect best practices or what researchers should do.

\subsection*{Freedom to choose the family of comparisons}

When using multiple testing procedures, researchers must define a family or collection of hypothesis tests to be corrected together, and exclude other hypothesis tests that may be conducted. Correction therefore occurs within families, not between families. Researchers are free to arbitrarily choose the family of tests, and although some conventions are followed, this is a often a cultural practice rather than a principled choice. As an example, suppose an experiment is designed as a 2-way ANOVA with two genotypes (wildtype and knock-out) and three doses of a compound (none, low, high). A standard ANOVA analysis yields three p-values: one for each main effect and another for the interaction. These three p-values are rarely corrected for multiple testing \cite{Cramer2015}, but they are a family of related tests since they use the same data, are part of the same statistical model, and use the same error term when calculating the p-values. Ryan argued in 1959 that these p-values should be corrected, but it has not become a standard practice \cite{Ryan1959}. However, researchers often correct for multiple testing when comparing levels within a factor, such as none versus low, and none versus high within the wildtype group. 

Oddly, the family of tests in the above experiment changes if it is a functional genomics study with many outcome variables, despite having an identical design. The family is usually a single comparison across all genes, so all $\sim$20,000 genes tested in the wild-type group comparing none versus low dose would be considered a family of tests. The p-value adjustment disregards the other treatment group and genotype. It may seem more rigorous to adjust 20,000 tests for a single comparison, as opposed to adjusting for all possible comparisons for one gene. But, doesn't adjusting for 20,000 genes times the number of group comparisons make sense as a strategy to limit the number of false positives across the entire experiment? Studies with more complex designs have multiple genotypes, treatment groups, time points and interactions between these, further complicating the determination of the family of tests \cite{Grabitz2018}.

Other reasonable families could be considered, such as all tests relevant to a research question. For example, does lesioning brain region X in mice affect their behaviour? Several cognitive, affective, and motor assays, as well as multiple readouts for each assay could be controlled as part of the same family. That is, the FPR could be controlled across all behaviours and readouts. Or, if the research question relates specifically to motor behaviours, then all the motor assays and the multiple readouts for each motor assay can be considered a family, and the FPR can be controlled for this family. It is rare for all the tests that relate to a research question to be grouped as a family, but it is a sensible way to control error rates. Another reasonable family consists of all the tests reported in a figure. The graphs in a multi-panel figure are often related and could be considered a family, and a researcher may wish to control this overall error rate. A more stringent approach is to adjust all the p-values reported in a paper, such that the probability of at least one false positive amongst all the reported hypothesis tests is 0.05. But what about tests reported in the supplementary material, or tests conducted but not published, or those published in another paper \cite{Rothman1990,Perneger1998}? 

My position is not that other families should be used, but that there are many sensible options and that current practices are arbitrary \cite{OKeefe2003}, and likely driven by software defaults, which then become cultural norms. Feise argues that families of tests are not only arbitrary, but also ambiguously and inconsistently defined \cite{Feise2002}. Hence, researchers may choose to group their hypothesis tests into many small families instead of fewer larger ones so that less stringent corrections are applied. This provides weaker control of the FPR and appears to be standard practice. Additionally, researchers may avoid reporting non-significant comparisons because a less stringent correction is required for the remaining hypotheses \cite{Althouse2016}. Such questionable research practices contribute to the reproducibility crisis and are difficult for peer reviewers and readers to detect.

\subsection*{Freedom to choose the error rate to control}

Once researchers define the family of tests to be corrected, they can choose from four main error rates to control. The first is the \textit{per-comparison error rate} (PCER), which is the error rate for one comparison and is equal to the significance threshold defined by the researcher, usually $\alpha = 0.05$. The PCER does not take into account the number of hypotheses tested, and therefore does not correct for multiple testing. This is the appropriate error rate for a single planned comparison.

The second and most common one is the \textit{familywise error rate} (FWER)\footnote{Sometimes the term \textit{experimentwise error rate} is used. We use the more general term familywise since an experiment is only one way of defining a family.}, which is the probability of having at least one false positive in a collection of tests, when the null hypothesis is true for all tests. This is the classic error rate and is calculated as FWER = $1 - (1 - \alpha)^m$, where $\alpha$ is the error rate for a single test and $m$ is the number of tests conducted. Thus if $\alpha = 0.05$ and $m = 15$, we expect $1 - (1 - 0.05)^{15} = 0.54$, or a 54\% chance of at least one false positive among the 15 tests. This example highlights how one of the oldest and simplest MCP works. The Bonferroni correction controls the FPR using a smaller $\alpha$ cutoff, defined as the $\alpha$ for a single test divided by the number of tests ($\alpha_\mathrm{Bonferroni} = 0.05/15 = 0.0033$). The FWER is then $1 - (1 - \alpha/m)^m = 1 - (1 - 0.0033/15)^{15} = 0.0488$, or approximately the target of 0.05, rounded to two decimal places. Alternatively, instead of using a new cutoff, the p-values can be adjusted upward, and for the Bonferroni correction the raw p-value is multiplied by the number of tests, so if the raw p-value is 0.0021, the adjusted value is $0.0021 \times 15 = 0.0315$. Many MCPs work in a similar way; a new cutoff is calculated from the number of tests (or the p-value adjusted), but the details vary for each method.

The third and most recent error rate is the \textit{false discovery rate} (FDR) \cite{Benjamini1995}. The FDR is the proportion of false positives amongst all the tests declared statistically significant. If 1000 tests are conducted and 200 are deemed significant at an FDR = 0.05 level, then the interpretation is that 200 $\times$ 0.05 = 10 are false positives and the other 190 are true positives or true discoveries. The FDR has weaker control over the FPR and is often used for high throughput and high dimensional ``omics'' experiments, which have few samples but many tests.

If controlling the FWER results in no or few significant p-values, researchers are free to move the goal posts and control the FDR instead, or they can choose the FDR from the start, thereby ensuring weaker error control. The FDR is now occasionally used instead of the more rigorous FWER in experiments with few hypothesis tests.

The fourth and least common error is the \textit{per-family error rate} (PFER) \cite{Ryan1959,Steinfatt1979,Klockars1994}, which is the expected rate at which false positives occur in the family of tests. It is defined as the number of expected errors out of $m$ tests, so if the error rate for a single test is $\alpha = 0.05$ and 2000 tests are conducted, the expected number of errors is $2000 * 0.05 = 100$. PFER is rarely selected since it is the most stringent of the four methods, resulting in fewer significant p-values. However, the popular Bonferroni correction controls both the PFER and FWER, so it is often used in practice. The Bonferroni correction is often criticised for being overly conservative, but it is only conservative for the FWER; it is well calibrated for the PFER \cite{Klockars1994,Frane2015}. If the goal is to minimise the number of false positives, then the PFER is recommended over the FWER; whereas the FWER is recommended if, given one false positive, subsequent false positives are less relevant \cite{Ryan1959,Frane2015}.

\subsection*{Freedom to choose the correction method}

After deciding the family of tests and the error rate, researchers have (too) many options for the method of correction: for the FWER, options include Bonferroni \cite{Bonferroni1936}, progressive Bonferroni \cite{Hewitt1997}, Holm \cite{Holm1979}, Scheffe \cite{Scheffe1959}, Sidak \cite{Sidak1967}, Tukey \cite{Tukey1949}, Student-Newman-Keuls \cite{Newman1939,Keuls1952}, Dunn \cite{Dunn1964}, Duncan \cite{Duncan1955}, Dunnett \cite{Dunnett1955} and so on; and for the FDR, options include Benjamini-Hochberg \cite{Benjamini1995}, Benjamini-Yekutieli \cite{Benjamini2005}, Q-value \cite{Storey2002,Storey2003a}, local FDR, positive FDR \cite{Storey2002,Storey2003a} marginal FDR \cite{Tsai2003}, conditional FDR \cite{Tsai2003}, empirical FDR \cite{Tsai2003}, Black Box FDR \cite{Tansey2018}, Beta-Uniform mixture \cite{Pounds2003}, and so on. Dozens of methods exist to correct for multiple comparisons and researchers are free to choose (subject to restrictions on the type of comparison), sometimes making their selection based on the results, and choosing the method that provides significant p-values for the key comparisons. Modern software makes it easy to click many options, examine the output, and report the procedure that gives the desired results (recommendations are available about preferred methods \cite{Day1989,Midway2020}).

More knowledgeable researchers can avail themselves of further options. Many older MCPs adjust all the hypotheses in a family together, such as the Bonferroni example above. However, researchers can order hypotheses using fixed-sequence, fallback, gatekeeping, and recycling procedures, where hypotheses earlier in the order are subject to fewer, if any, corrections. \cite{Bauer1998,Westfall2001,Wiens2003,Dmitrienko2003,Huque2008,Li2008,Huque2018}. As an example, in a study with a low dose, high dose, and control group, the control versus high dose comparison would be tested first, and if this is significant, only then would the control versus low dose group be tested, with no penalty for having conducted the first test. As a result, by declaring that a ``fixed-sequence procedure'' was used, researchers can avoid any real correction for multiple testing. These methods are rarely used outside of clinical trials, but only because they are unfamiliar to most researchers and unavailable in popular statistical software. By continually developing new ways to trade-off false positives and power, methods for adjusting p-values have become more complex than the statistical models themselves \cite{Bretz2011a}.

Stepping back from the purpose and justification of MCPs and the mathematics underpinning them, we see that they are merely rank-preserving transformations of p-values. In other words, these methods monotonically map p-values -- numbers between 0 and 1 -- to a new set of numbers, also between 0 and 1. Graphing p-values on the $x$-axis and adjusted p-valued on the $y$-axis plots a convex curve, and the various methods only ``bend the curve'' differently. The usefulness of controlling false positives by such remapping is open to debate. Steinfatt proposed a simple alternative in 1979 that has been forgotten but which would be suitable for many experiments \cite{Steinfatt1979}. He suggested dividing the expected number of false positives by the observed number of significant results to obtain what he called the Alpha Percentage ($\alpha$\%). For example, taking $\alpha = 0.05$ and assuming 1000 tests are conducted, the expected number of false positives is $1000 * 0.05 = 50$, and if 68 significant results are observed, $\alpha$\% = Expected/Observed $\times 100 = 50/68 \times 100 \approx 74$\%. Hence, about a quarter of significant results are likely false positives. This percentage will be reduced as the number of observed significant p-values increases, and hence a lower $\alpha$\% is better. Readers can then interpret the unadjusted p-values in light of this number. Steinfatt also suggested calculating $\alpha$\% for several $\alpha$ values to better understand how the number of excess significant p-values depends on $\alpha$.

\subsection*{Procedures that provide poor control of the FPR are often used}

Since there are so many correction procedures, it is not surprising that some control error rates less effectively than others, and that researchers will sometimes use them. At the extreme end, Fisher's Least Significant Difference (LSD) provides no correction for multiple testing. The LSD is a rule-of-thumb to ``only do post-hoc comparisons if the overall ANOVA F-test is significant''. This if-then rule is rarely used, and even when it is, control of the FPR is only adequate when there are three groups or comparisons \cite{Meier2006}. Furthermore, the rule-of-thumb is inappropriate for experiments with positive controls because the overall F-test will always be significant (if not, there is a problem with the experiment). Despite these problems, researchers frequently report using Fisher's LSD. The methods and results sections of journals from the Public Library of Science, BioMed Central, and Nature Publishing Group were searched on 22 June 2021 and approximately 3700 papers reported using this method.

Other procedures such as the Newman-Keuls or Student-Newman-Keuls (SNK) provide only weak control of the FPR. It's better than nothing, but the true FPR is higher than the advertised 5\% level, and error control gets worse as more comparisons are made. SNK is commonly used despite warnings from statistical software manufacturers: ``We offer the Newman-Keuls test for historical reasons\ldots{} but we suggest you avoid it because it does not maintain the family-wise error rate at the specified level''\footnote{\url{https://www.graphpad.com/guides/prism/7/statistics/stat_options_tab_three-way_anova.htm}}. Over 10,000 papers reported using this method from the above publications.

With big effect sizes and few comparisons, many papers using Fisher's LSD or SNK would have come to the same conclusion if a better method had been used. However, since a disproportionate number of published p-values are suspiciously close to 0.05 \cite{Ridley2007,Masicampo2012,Leggett2013,Winter2015,Krawczyk2015,Zwet2021}, more stringent control of the FPR will no doubt change the conclusion for many.

\subsection*{Exploratory analyses are uncontrolled}

Prior to a formal analysis, statisticians recommend plotting and graphically exploring the data to better understand the data and check for outliers, clusters, unusual or impossible values, etc. \cite{Cleveland1993,Cleveland1994,Krause2012,Lazic2016}. Visual inspection of the data is strongly recommended; but, unanticipated patterns and relationships are often found, which lead to new hypotheses. Unfortunately, the same data cannot be used to both generate and test a hypothesis, at least if control of the FPR matters. This process has been called ``HARKing'' -- Hypothesising After the Results are Known \cite{Kerr1998}.

All MCPs take the number of tests into account, but this number is unknown when data have been examined for patterns because it includes all the tests that \textit{might} have been done if the pattern was interesting enough \cite{Groot2014}. Even though only one test was conducted, it needs to be corrected for other potential tests that might have been done, which is impossible to determine. Hence, there is no way to adjust p-values to account for data-driven hypotheses when testing for discovered relationships.

\subsection*{Alternative analyses are uncontrolled}

Researchers often try several models before settling on a final analysis. This includes adding or removing covariates such as age or sex, transforming the outcome to achieve normality or to stabilise variances, modifying likelihood and link functions for generalised linear models, and so on. The most appropriate model cannot always be determined when planning an experiment, so it makes sense to try alternate models until a suitable one is found. But researchers can then pick a model that gives the desired results and supports their theory. Simmons and colleagues showed how easy it is to get significant results with such alternate analyses \cite{Simmons2011}. Again, often there is no clearly defined family of tests and therefore no way to adjust for multiple alternative analyses.

\subsection*{Multiple analyses as data accumulate are uncontrolled}

In many settings the data accumulate over time and are analysed before the experiment is complete. Examples include clinical samples, patients, or other human subjects that are recruited into an experiment; cell culture experiments where the whole procedure or protocol is repeated several times; or animal experiments that are run on each litter of animals as they are born. Multiple ``looks'' increases the FPR because there are more chances for results to be significant, especially when an experiment is terminated early because the desired result was found. Although there are ways to control the FPR in these situations \cite{Berry2010,Schonbrodt2017}, they are rare outside of clinical trials.

\subsection*{Analyses of the same data by multiple research groups are uncontrolled}

An underappreciated situation that increases the FPR is when multiple individuals or research groups analyse the same data \cite{Thompson2020}. This is becoming more common with data being deposited in repositories and consortia pooling data, but members analyse them separately. Even if each research group carefully controls the FPR, the same data is analysed in so many ways by so many people that every false positive result will likely be found and reported. For example, there are over 2000 primary publications in PubMed using data from the Alzheimer's Disease Neuroimaging Initiative (ADNI) and nearly 15,000 primary publications using the National Health and Nutrition Examination Survey (NHANES) data. Neither example has a single fixed dataset, but multiple ones that have accumulated over time, so they suffer both from ``multiple looks'' and multiple analysts, making it hard to define a family of tests. Grootswagers and Robinson report a similar problem in computational cognitive neuroscience \cite{Grootswagers2021}.

\subsection*{Suggestions from people in authority are uncontrolled}

Finally, it is not uncommon for peer reviewers, editors, collaborators, clients, PhD/postdoc supervisors, or other senior individuals to suggest alternative analyses, tests, transformations, and removal of outliers -- often with the best of intentions. Even if a researcher maintained strict control of the FPR, the additional requests often come after the analysis is done and can increase the FPR. These ``requests'' may be hard to ignore by junior researchers and are a source of friction for professional data analysts. Although the suggestion from a more senior and knowledgeable person may be a good idea, it still weakens strict error control.

\section*{Solutions}

The distance between the true and nominal FPR can be reduced in many ways, but rarely will they be made equal. A simple solution is to avoid MCPs that don't control FPR well, such as Fisher's LSD and Newman-Keuls methods. Researchers can switch to other methods and journals can discourage their use. Tukey's or Holm's method can be substituted in all cases, or more recent ``simultaneous tests'' or ``multiple marginal models'' that account for correlations between parameters \cite{Bretz2011,Pipper2011,Pallmann2018,Vogel2020}. If corrections for multiple tests are not used, this should be clearly stated.

FPRs get inflated when researchers have too much freedom to choose the family, error rate, and correction method. Defining the analytical decisions before the experiment is conducted limits the choices. A pre-registered analysis plan specifies the options and prevents data-driven decisions increasing the FPR. Nevertheless, the most liberal options can be chosen beforehand, which mitigates the effectiveness of this approach. In addition, vague language like ``\ldots the main analysis will be followed by post-hoc tests as appropriate'' is pointless and no better than a standard non-preregistered study. Constraints are best suited for confirmatory experiments where there's a clear hypothesis, a primary outcome variable, and prior information to help define a statistical model. However, since confirmatory experiments tend to have few hypotheses, the risk of false positives is lower, and so MCPs likely work best where they are needed the least.

Inflated FPRs also result from uncontrolled analyses: exploratory, alternative, as data accumulate, by multiple groups, and from suggestions from others. In these cases it is difficult or impossible to define the family of tests, and so the standard MCPs cannot be used to control the FPR. One solution is shrinkage, also known as regularisation or partial pooling, and is described below.

\subsection*{Shrink parameter estimates}

There's a debate over whether it's better to estimate parameters (like effect sizes) or to test hypotheses \cite{Cumming2013,Morey2014}. The methods are complementary and many studies report both p-values and effect sizes with confidence intervals. Here, I am not arguing that parameter estimation should replace hypothesis testing, but that parameter estimation helps solve multiple testing. The problem is not that some p-values are too small, but that some effect sizes are too big. Therefore the solution is to shrink parameters closer to zero or the null value \cite{Zwet2021}. This approach recognises that \textit{p-values are not the problem, it's the effect sizes}. Adjusting p-values is like treating the symptoms of a disease instead of the cause. Furthermore, adjusted p-values do not fix the visual impression of exaggerated effects when the data are plotted.

What causes a small p-value when there's no true effect or association, assuming that the experiment is carefully controlled and unbiased? The p-value is based on a 1:1 mapping from a test statistic (along with the degrees of freedom), so the test statistic must be too big (Fig. 1). What causes a large test statistic? Even when there is no real difference or association, sometimes a large effect size is observed due to sampling error or some other confounding. A second reason is that even if the effect size is small or moderate, it may be estimated with high precision. Precision is a function of the sample size (N) and the variability of the data (SD), specifically the within-group variability. Sometimes the SD is underestimated so that even a small effect size can lead to a large test statistic and hence a significant p-value. Would it not make more sense to fix the root cause instead of a downstream effect?

\begin{figure}
\centering
\includegraphics[scale=1.2]{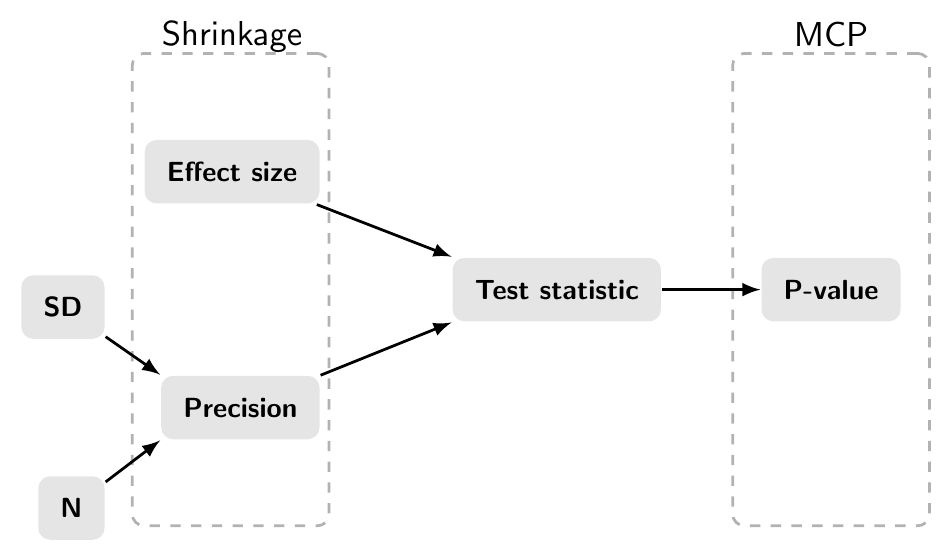}
\caption{What causes a small p-value? Ultimately, a large effect size or a small within-group standard deviation (SD). Shrinkage methods typically adjust both the effect size and the precision, which is a function of the SD and sample size (N). Multiple comparison procedures (MCPs) adjust only p-values.}
\end{figure}

Shrinking parameters follows this logic: say we're estimating the difference between the means of two groups, call this $\theta$. Even if $\theta = 0$, our experiment will rarely return a value of exactly 0, sometimes it will be greater than 0 and sometimes less than 0. Sometimes $\theta$ will be far from 0 because of sampling error and we will incorrectly conclude that $\theta \neq 0$. Knowing this, we can take the observed value of $\theta$ and pull or shrink it closer to 0, so that we don't get large values when there's no effect, thereby reducing the number of false positives. Although shrinking might reduce statistical power since true effects will also be shrunk towards 0, often the reduction in false positives will offset that loss. This counter-intuitive result, where the best estimate of $\theta$ is not the measured or directly calculated value of $\theta$, is known as Stein's Paradox \cite{Efron1977,Samworth2012}.

Figure 2 shows shrinkage with simulated data. Data were simulated for two groups with ten samples each, and for 100 genes. The mean difference between the groups for each gene was compared with an independent samples t-test. Two data sets were simulated, one without group differences and one with five genes differing between groups. Figure 1A shows the estimates (mean difference) and standard errors for the 100 genes with no true effect. Figure 1B shows the shrunk estimates, which have all been pulled to zero with little uncertainty, indicating that the model is confident that no effects exist. Figure 1C shows estimates for the data set with five genes having different expression levels (all with negative values). Figure 1D shows the shrunk estimates, and while most genes have been shrunken to zero, four remain distinct. Note how shrinking corrects the inflated effect sizes.

\begin{figure}
\centering
\includegraphics[scale=0.6]{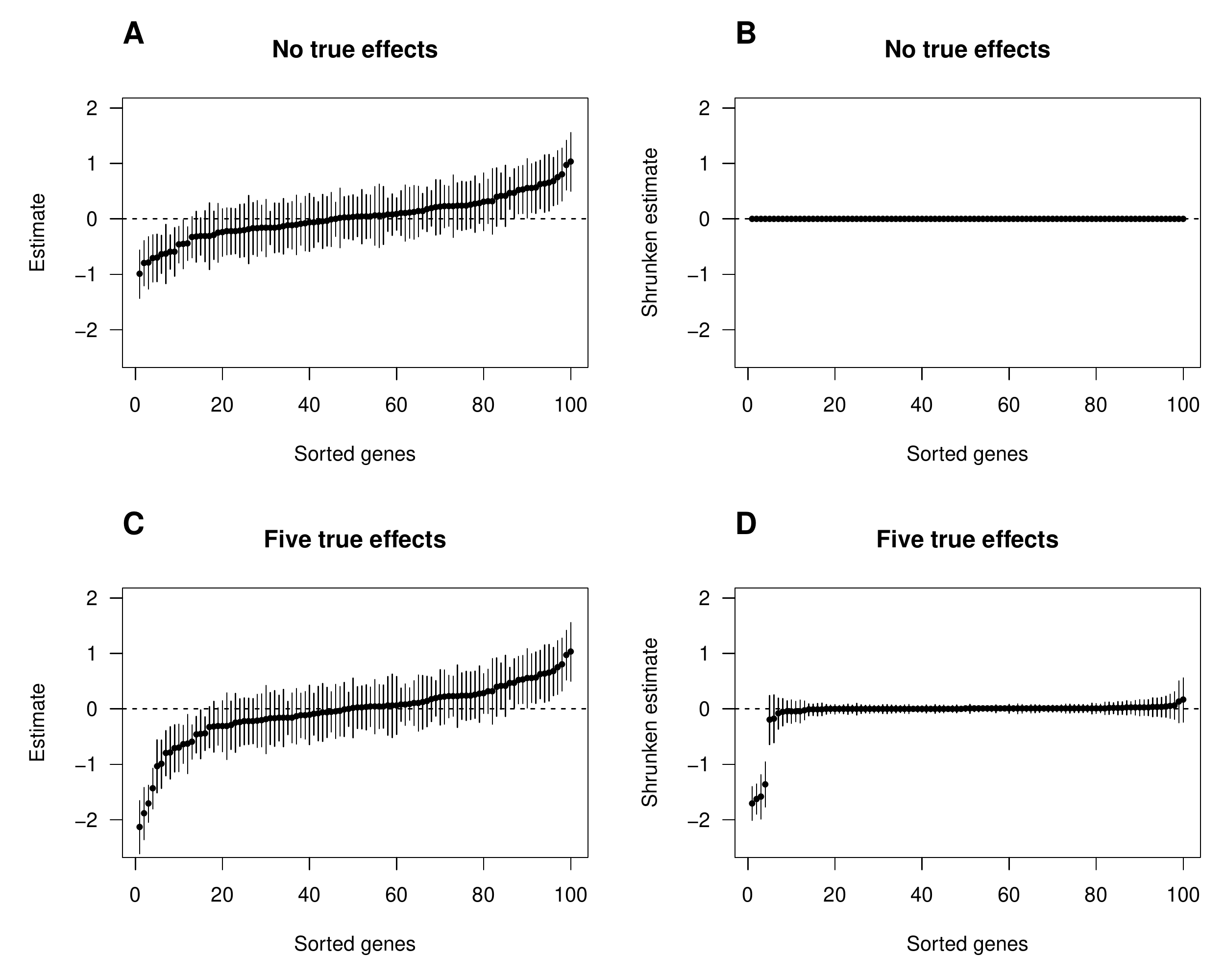}
\caption{Shrinking parameters. Estimates (mean difference) and standard errors for the data with no true effects (A), and the shrunken estimates (B), which have all been shrunk to 0. Estimates and standard errors for the data with five true negative effects (C), and the shrunken estimates (D), of which four remain strongly negative.}
\end{figure}

Parameter estimates can be shrunk in three ways (Table 1). The first is to take a Bayesian approach and place informative priors on the parameter $\theta$ \cite{Greenland1992,Greenland2015}. For example, when comparing several groups to a control, the parameters representing the difference between group means -- call them $\theta_i$ for $i=1, \dots, p$ comparisons -- could have independent Normal(0, $\sigma$) priors (see McElreath for an introduction to Bayesian methods \cite{McElreath2016}). The observed mean differences ($\theta_i$) would shrink towards zero, with the amount controlled by $\sigma$ (a small $\sigma$ provides more shrinkage). This is a general approach but the control of the FPR critically depends on $\sigma$. Often there is no theory to select a suitable value, but simulations can help. Data are simulated from hypothetical experiments with no effects present, a value for $\sigma$ is chosen, and the data analysed as they will be in the real experiment. Repeat this for many simulated datasets and count the number of false positives. If this number is greater than a desired threshold (e.g. 5\%), then $\sigma$ needs to be smaller. Repeat this procedure until a suitable value of $\sigma$ is found, and the sample size can also be calculated to ensure that statistical power is adequate. This approach works best when there are few tests, and no connection between them; for example, when comparing several clinical outcomes between a group of patients and controls, and each outcome is analysed separately.

\begin{table}[ht]
  \centering
  \caption{Three approaches to shrinking parameter estimates.}
  \begin{tabular}{ll}
    \hline
    \textbf{Method} & \textbf{When to use} \\
    \hline
    Informative priors & Few tests, no relationship between them \\
    Multilevel model & Moderate number of tests, same outcome  \\
    Two-step &  Many tests, same outcome \\
    \hline
  \end{tabular}
\end{table}

The second approach uses a Bayesian or empirical Bayesian multilevel/hierarchical model \cite{Gelman2012}. Here, we assume the $\theta_i$ parameters come from a common distribution with a mean of zero and a standard deviation of $\sigma$. Unlike the first approach, $\sigma$ is estimated from the data, and hence the amount of shrinkage is data-dependent: with little variation between the group means, $\sigma$ will be small and the shrinkage will be greater. This approach only works if outcomes are measured on the same scale since all the parameters are a part of one big model. An example is the data shown in Figure 2, where the expression of many genes is compared between groups. A drawback of this approach is that with few groups or comparisons, $\sigma$ will be hard to estimate, and so it's best used with a moderate number of tests. 

Empirical Bayes methods are especially useful since there are no options to select once the model structure is defined. Shrinkage is data-dependent, with the amount of shrinkage depending on (1) the ratio of variability within and between groups, (2) the number of parameters, and (3) the number of samples within each group. A key feature is that the rank order of effects can change, unlike with p-value adjustments, where the ordering of adjusted p-values stays the same.

The third approach is a two-step method. First, summary statistics for the comparisons are computed, such as effect sizes $\theta_i$ and their standard errors. Then, empirical Bayesian methods shrink the effect sizes \cite{Thomas1985,Greenland1991,Stephens2017}. This approach is especially useful for ``omics'' and high-dimensional biology experiments with many hypotheses, as in Figure 2 \cite{Stephens2017}. Another approach developed by van Zwet and Gelman uses the signal-to-noise ratio of similar published studies to estimate the prior distribution, which is then used to shrink the parameters \cite{Zwet2021a}.

The main drawback with shrinking parameters to control false positives is that the FPR can't be determined precisely before experiments are performed, although it can be estimated with simulation studies. The traditional approach provides theoretical guarantees (e.g. that the FWER for $n$ tests will be $\leq \alpha$ when using $\alpha/n$ as the significance threshold), which rarely applies to real experiments, as argued above. Hence, an approximate error rate that's actually achieved is better than an exact theoretical error rate that doesn't apply. A second drawback is that these methods are not implemented in popular software packages and so require more work than ticking boxes, although they are available in R for omics experiments \cite{Stephens2017} and correlation matrices \cite{Schaefer2005}.

Several advantages outweigh these drawbacks: results are less dependent on the number of tests, adjustments can be made when the number of tests is not defined, and options are fewer, so it's harder for researchers to select methods that give desired results. Researchers must choose a prior for the Bayesian methods, but these can also be specified in advance. Shrinkage is also a more direct solution since exaggerated effects are brought closer to the true effects, on average.

\section*{Discussion}
The FPR is commonly viewed as a single well-defined entity, but the arguments above make this idea untenable. First, many \textit{error rates} exist -- per-comparison, familywise, false discovery, and per-family -- so by definition there cannot be a generic ``false positive rate'' that a researcher can control. Multiple comparison procedures critically depend on the number of tests included in the family, which is often hard to define, such as when the data accumulate over time, when several people analyse the same data, when multiple analyses are tried, and for exploratory studies. Berry called these silent or hidden multiplicities \cite{Berry2007,Berry2012}. Consequently, adjusted p-values are impossible to calculate and therefore meaningless -- just like you cannot solve for $y$ when $x$ is unknown in $y = 12/x$. It might be better to acknowledge that $y$ is undefined instead of making up a fictitious value for $x$ to calculate $y$. In cases where the family can be clearly defined, it's usually arbitrary, and other sensible families could have been chosen, leading to different results. Finally, the bewildering number of procedures available means that different results can be obtained for the same supposed error rate.

When studies assess the performance of MCPs, they always take the family as unambiguously given, but since this is an ill-defined entity, performance of MCPs in the real world is questionable. While there's no direct proof, I speculate that one reason for the ``reproducibility crisis'' is that the classic MCP approach doesn't work as well as it should. It is an ill-posed problem because researchers are trying to control something that cannot be defined, measured, and may not even exist. This means that the usual fixes such as improving researcher incentives and training will have little impact on this aspect of reproducibility. Although, limiting researcher degrees of freedom by defining the analysis upfront and avoiding methods with poor performance can help.

The classic MCP approach assumes that effects can be logically separated into those that exist and those that do not, which is a simplistic view of the world \cite{Gelman2015}. Like p-values, MCPs have a social function: they let researchers make scientific claims. For this reason, p-values are not going away, despite calls to reform, ban, or abandon their use (see Volume 73 of \textit{The American Statistician} \cite{Wasserstein2019}). In most fields, a researcher is allowed to claim there's an effect or an association if the p-value or adjusted p-value is less than 0.05. This makes MCPs necessary, the argument goes, since the likelihood of false claims increases when many claims are made. We should keep false claims under control, but is adjusting p-values really the best way? I argue the best quantity to adjust is effect sizes and their precision, not p-values. Shrinkage methods are rarely used to control the FPR in most fields, and so little is known about their performance in the real world, but it is worth exploring further. For example, how would these methods work if subsets of a large dataset were analysed separately, mimicking the situation where multiple people analyse parts of the data? MCPs are expected to have weaker control for the subsets since the family of tests is smaller, and hence the FPR will likely be inflated.

Since estimated effect sizes are too large, shrinking parameters is fundamentally the right approach to make fewer false scientific claims. Since p-values are calculated from the effect sizes, the argument is to move the adjustment upstream. If effect sizes are improved, so are p-values; much like treating the disease often improves the symptoms. However, shrinkage methods are harder to apply because they require more knowledge than ticking a box. In addition, the best approach will depend on the experimental design. Some experiments are simple with few comparisons, like a clinical trial that compares a low and high dose of a compound against a placebo. Other experiments have simple designs (e.g. healthy versus disease) but with thousands of outcomes, such as gene or protein expression values. Others have a single outcome but many comparisons, such as a screen testing thousands of drugs. Finally, some experiments run several assays on the same samples or subjects, each with various outcomes and factors (e.g. a typical neuroscience experiment with several behavioural assays and several outcomes per assay). The best way to shrink estimates will probably vary between designs, and we don't have any data comparing the options. Shrinking estimates gives up the theoretical guarantees of the MCP approach, but since these guarantees do not apply in many real world settings, little is lost.
 
The paper encourages scientists to use shrinkage methods in their research, perhaps along with standard ones, and to compare the two. Eventually, scientists will be able to compare the relative merits of this approach on the types of data they routinely generate. We also need quantitative researchers to assess these methods using simulations to understand when they work well, when they fail, and if there are sensible default options. This would be more valuable than developing yet another multiple comparison procedure. In the end, we shouldn't assume effects either exist or don't exist, and try to reduce false positives; instead, we should minimise the discrepancy between true and estimated effects, and wrong conclusions will be automatically controlled.

\bibliography{manuscript_v2}
\end{document}